\documentclass[12pt]{article}
\usepackage{graphicx}

\begin{document}
\title{Form factor of $\pi^0\rightarrow\gamma\gamma^*$}
\author{Bing An Li\\
Department of Physics and Astronomy, University of Kentucky\\
Lexington, KY 40506, USA}

\maketitle
\begin{abstract}
An effective chiral theory of large $N_C$ QCD of mesons has been applied to
study the form factor of $\pi\gamma\gamma^*$. Besides the poles of vector mesons
an intrinsic form factor is found. The slope of the form factor is
predicted. The effect of the current quark masses on the decay rate
is calculated.
There is no adjustable parameter in this study.
\end{abstract}
\newpage
It is well know that the decay amplitude of $\pi^0\rightarrow\gamma\gamma$ has
been predicted by Adler-Bell-Jackiw[1], the famous triangle anomaly. This decay
is a very important process in both particle physics and field theory. The form
factor of $\pi\gamma\gamma^*$ is associated with both anomaly and Vector Meson
Dominance(VMD). The measurements of the form factor has lasted for a long time[2,3].
In the timelike region of the slope of the form factor of
$\pi^0\rightarrow\gamma e^+ e^-$
\[F(q^2)=1+a\frac{m^2_{e^+ e^-}}{m^2_{\pi^0}}\]
has been measured[3] and the value of the slope a is in a wide range
\[-0.24\leq a \leq 0.12.\]
Recently,
the PrimEx Collaboration of JLab proposes to do direct
precision measurements of the slope a
in a range of $0.001GeV^2 \leq q^2 \leq 0.5 GeV^2$[4]. On the
other hand, the form factor of $\pi^0\gamma\gamma^*$ has been studied by various
theoretical approaches[5].
 
In Ref.[6] we have proposed an effective chiral theory of large $N_C$ QCD of
pseudoscalar, vector, and axial-vector mesons.
This theory is based on t'Hooft's
large $N_C$ expansion of QCD[7]. As interpolating fields, meson fields
are
coupled to quarks and they are not independent degrees of freedom.
In the limit $m_q\rightarrow 0$, the effective theory is chiral symmetric and
has dynamical chiral
symmetry breaking. The Lagrangian is expressed as[6]
\begin{eqnarray}
{\cal L}=\bar{\psi}(x)(i\gamma\cdot\partial+\gamma\cdot v
+\gamma\cdot a\gamma_{5}+eQ\gamma\cdot A
-mu(x))\psi(x)-\bar{\psi(x)}M\psi(x)\nonumber \\
+{1\over 2}m^{2}_{0}(\rho^{\mu}_{i}\rho_{\mu i}+
\omega^{\mu}\omega_{\mu}
+a^{\mu}_{i}a_{\mu i}
+f^{\mu}f_{\mu})
\end{eqnarray}
where M is the quark mass matrix
\[\left(\begin{array}{c}
         m_{u}\hspace{0.5cm}0\\
         0\hspace{0.5cm}m_{d}
        \end{array}  \right ),\]
\(v_{\mu}=\tau_{i}\rho^{i}_{\mu}
+\omega_{\mu}\)
,
\(a_{\mu}=\tau_{i}a^{i}_{\mu}
+f_{\mu}\)
,
and \(u=exp\{i\gamma_{5}(\tau_{i}\pi_{i}+
\eta)\}\).
The kinetic terms of mesons are generated by quark loops. The Lagrangian of mesons is
obtained by integrate the quark fields out.
After renormalization physical
meson fields, pion decay constant $f_\pi$, and a universal coupling
constant g are defined. $f_\pi$ and g are two inputs. g is
determined to be 0.395 by fitting the decay rate of $\rho\rightarrow ee^+$.
$N_C$ expansion is revealed from this theory.
The tree diagrams are at leading order of $N_C$
expansion and loop diagrams of mesons are at higher orders. Adler-Bell-Jakiw(ABJ) and
Wess-Zumino-Witten(WZW)[8] anomaly is the imaginary part of the Lagrangian of mesons[6].  	
VMD is a natural result of this theory[6]. We have applied this theory to study various
meson physics and the theory is phenomenological successful[9,10].
 
The form factors of charged pion and kaons have been studied by this theory[9]. It has
found that
besides the poles of vector mesons, there are
additional form factors which are called intrinsic form factors caused by quark
loop effects. For example, the form factor of charged pion is determined to be[9]
\begin{eqnarray}
F_\pi (q^2) & = & f_{\rho \pi \pi }(q^2)
\frac{-m_\rho ^2+i\sqrt{q^2}\Gamma _\rho
(q^2)}{q^2-m_\rho ^2+i\sqrt{q^2}\Gamma _\rho (q^2)},
\end{eqnarray}
where $\Gamma _\rho(q^2)$ is the decay width of $\rho$[9]. In the space-like region
\(\Gamma_\rho=0\). Besides a $\rho$ pole, there is an intrinsic form factor
$f_{\rho\pi\pi}(q^2)$
\begin{equation}
f_{\rho\pi\pi}(q^2)=1+\frac{q^2}{2\pi^2 f^2_\pi}\{(1-{2c\over g})^2
-4\pi^2 c^2\},
\end{equation}
\[c=\frac{f^2_\pi}{2gm^2_\rho}.\]
The effects of the intrinsic form factor are:
a larger radius of charged pion is revealed from Eq.(2); in timelike
region the $|F_\pi|$
decreases faster and in spacelike region $F_\pi$ decreases slower than the
$\rho$ pole
only. These results agree well with data.
 
In Ref.[6] we have studied the form
factor of $\pi^0\gamma\gamma^*$. However, only the poles of vector mesons are
taken
into account. In this paper the intrinsic form factor of $\pi^0\gamma\gamma^*$
is
studied and the slope is predicted. The effect of current quark mass
on the decay rate
of
$\pi^0\rightarrow\gamma\gamma$ is investigated too. In these studies there is
no adjustable parameter.
 
There are four processes which contribute to the form factor of
$\pi^0\gamma\gamma^*$.
The four processes are shown in Fig.1.
In these processes there are four vertices: ${\cal L}^{\pi^0\omega\rho}$,
${\cal L}^{\gamma\rho}$, ${\cal L}^{\gamma\omega}$, and
${\cal L}^{\pi^0\gamma\gamma}$.
In Ref.[6] up to the fourth order in derivatives
these vertices have been derived from Eq.(1). ABJ anomaly is obtained
\begin{equation}
{\cal L}_{\pi^0 \rightarrow\gamma\gamma}=-\frac{\alpha}{\pi f_\pi}
\varepsilon^{\mu\nu\lambda\beta}\pi^0 \partial_\mu A_\nu \partial_\lambda
A_\beta
\end{equation}
and it is found that
the form factor of $\pi^0\rightarrow\gamma\gamma^*$ is the poles of $\rho$ and
$\omega$ mesons. However, like the form factor of charged pion(2), because of quark loop effects besides the poles of
vector mesons we should expect
an additional form factor for $\pi^0\rightarrow\gamma\gamma^*$ too.
In order to
investigate this intrinsic form factor
we need to derive the four
vertices
to sixth order in derivatives.
From Eq.(1) related interaction Lagrangian is obtained
\begin{equation}
{\cal L}=\bar{\psi}\{{1\over g}\gamma\cdot
(\tau_3\rho^0+\omega)-i{2m\over f_\pi}\tau_3
\gamma_5 \pi^0
-{2\over f_\pi}{c\over g}\tau_3 \gamma_\mu\gamma_5 \partial^\mu\pi^0+eQ\gamma_\mu A^\mu\}\psi.
\end{equation}
The term $-{c\over g}\partial_\mu\pi^0$ is obtained from the transformation[6]
\[a^0_\mu\rightarrow a^0_\mu-{2\over f_\pi}{c\over g}\partial_\mu\pi^0\]
which is used to cancel the mixing between $a_\mu$ and $\pi$ fields.
In Ref.[6] the
formalism of the path integral has been used to calculate
the quark loop diagrams up
to the fourth order in derivatives. In this paper we use the
interaction Lagrangian(5)
to calculate quark loop diagrams to find the meson vertices up to
the sixth order in derivatives. We obtain
\begin{eqnarray}
\lefteqn{{\cal L}_{\rho\gamma}=-{e\over4}g(\partial_\mu
A_\nu-\partial_\nu A_\mu)
\{1-{1\over10\pi^2 g^2}{\partial^2\over m^2}\}(\partial^\mu \rho^{0\nu}
-\partial_\nu
\rho^{0\mu}),}\\
&&
{\cal L}_{\omega\gamma}=-{e\over12}g(\partial_\mu A_\nu-\partial_\nu A_\mu)
\{1-{1\over10\pi^2 g^2}{\partial^2\over m^2}\}(\partial^\mu \omega^{\nu}
-\partial^\nu
\omega^{\mu}),\\
&&{\cal L}_{\pi^0\omega\rho}=-{3\over \pi^2 g^2 f_\pi}\pi^0
\varepsilon^{\mu\nu\lambda\beta}
\partial_\mu\rho^0_\nu\partial_\lambda\omega_\beta\{1+{1\over12m^2}
(1-{2c\over g})(
k^2_1+k^2_2+p^2)\},
\end{eqnarray}
where $k^2_{1,2}$ and $p^2$ are momentum(in fact, it is -$\partial^2$) of $\rho$,
$\omega$, and $\pi^0$ mesons
respectively. It is interesting to notice that in Eq.(5) there is a coupling between
$\partial_\mu \pi^0$ and the axial-vector current
$\bar{\psi}\tau_3\gamma_\mu\gamma_5\psi$.
The
term ${1\over12m^2}(-{2c\over g})(k^2_1+k^2_2+p^2)$ is obtained from this coupling.
Because of the strong anomaly[10] this term is not zero in the chiral limit.
 
In deriving Eqs.(6-8) the equation[6] with a cut off
\begin{equation}
\frac{N_C}{(4\pi)^2}{D\over4}m^2\int d^Dp\frac{1}{(p^2+m^2)^2}={F^2\over 16}
\end{equation}
has been used. Following equations[6] are useful in this paper
\begin{eqnarray}
\lefteqn{F^2(1-{2c\over g})=f^2_\pi,}\\
&&m^2={1\over6}{F^2\over g^2}.
\end{eqnarray}
Using the substitutions
\begin{equation}
\rho^0_\mu\rightarrow {1\over2}egA_\mu,\;\;\;\omega_\mu\rightarrow{1\over6}egA_\mu
\end{equation}
in Eq.(8), we obtain
\begin{eqnarray}
\lefteqn{{\cal L}_{\pi^0\gamma\gamma}=-\frac{e^2}{4\pi^2 f_\pi}
\{1+{g^2\over2f^2_\pi}(1-{2c\over g})^2
(k^2_1+k^2_2+p^2)\}\pi^0\varepsilon^{\mu\nu\lambda\beta}\partial_\mu A_\nu
\partial_\lambda A_\beta,}\\
&&{\cal L}_{\pi^0\rho\gamma}=-\frac{e}{2g\pi^2 f_\pi}
\{1+\frac{g^2}{2f^2_\pi}(1-{2c\over g})^2 (k^2_1+k^2_2+p^2)\}
\pi^0\varepsilon^{\mu\nu\lambda\beta}\partial_\mu A_\nu
\partial_\lambda \rho_\beta,\\
&&{\cal L}_{\pi^0\omega\gamma}=-\frac{3e}{2g\pi^2 f_\pi}
\{1+\frac{g^2}{2f^2_\pi}(1-{2c\over g})^2(k^2_1+k^2_2+p^2)\}
\pi^0\varepsilon^{\mu\nu\lambda\beta}\partial_\mu A_\nu
\partial_\lambda \omega_\beta.
\end{eqnarray}
 
Using Eqs.(6-8,13-15), the amplitudes of the processes of Fig.1 are calculated in the
chiral limit
\begin{equation}
<\gamma_1\gamma_2|S|\pi^0>=-i(2\pi)^4\delta^4(p-k_1-k_2)\frac{1}
{\sqrt{8m_\pi\omega_1\omega_2}}
\varepsilon^{\mu\nu\lambda\beta}\epsilon_\mu(1)\epsilon_\nu(2)k_{1\lambda}k_{2\beta}
\frac{2\alpha}{\pi f_\pi}F,
\end{equation}
where
\begin{eqnarray}
F& = &\{1+\frac{g^2}{2f^2_\pi}(1-{2c\over g})^2(k^2_1+k^2_2)\}
\{1-{1\over2}\frac{k^2_1}{k^2_1-m^2_\rho+i\sqrt{k^2_1}\Gamma_\rho(k^2_1)}
\nonumber \\
&&-{1\over2}\frac{k^2_2}{k^2_2-m^2_\rho+i\sqrt{k^2_2}\Gamma_\rho(k^2_2)}
-{1\over2}\frac{k^2_1}{k^2_1-m^2_\omega+i\sqrt{k^2_1}\Gamma_\omega(k^2_1)}
\nonumber \\
&&-{1\over2}\frac{k^2_2}{k^2_2-m^2_\omega+i\sqrt{k^2_2}\Gamma_\omega(k^2_2)}\},
\end{eqnarray}
where $\Gamma_\rho$ and $\Gamma_\omega$ are the total decay width of $\rho$ and $\omega$
meson respectively. The diagram Fig.1(d) is at higher order in derivatives. Therefore,
the contribution of this diagram is not included in Eq.(17).
In spacelike region they are zero. In timelike region[9]
\begin{eqnarray}
\Gamma _\rho (q^2) & = & \Gamma _{\rho ^0\rightarrow \pi ^{+}\pi
^{-}}(q^2)+\Gamma _{\rho ^0\rightarrow K\overline{K}}(q^2),
\nonumber \\ \Gamma _{\rho ^0\rightarrow \pi ^{+}\pi ^{-}}(q^2) &
= & \frac{f_{\rho \pi \pi }^2(q^2)\sqrt{q^2}}{12\pi g^2
}(1-\frac{4m_{\pi ^{+}}^2}{q^2})^{{3\over2}}\theta(q^2>4m_{\pi ^{+}}^2),
\nonumber
\\ \Gamma _{\rho ^0\rightarrow K\bar{K}}(q^2) & = & \frac{f_{\rho
\pi \pi }^2(q^2)\sqrt{q^2}}{48\pi g^2
}(1-\frac{4m_{K^{+}}^2}{q^2})^{{3\over2}}\theta(q^2>4m_{K^{+}}^2)
\nonumber \\ &&+\frac{f_{\rho \pi \pi }^2(q^2)\sqrt{q^2}}{48\pi
g^2}(1-\frac{4m_{K^0}^2}{q^2})^{{3\over2}}\theta (q^2>4m_{K^{0}}^2).
\end{eqnarray}
At \(q^2=m^2_\rho\), \(\Gamma_\rho=142MeV\).
There are
other channels for higher $q^2$. In Ref.[6,10] $\Gamma_\omega$ is calculated.
Put one of the two photons on mass shell, $k^2_2=0$, we obtain
the form factor of
$\pi^0\gamma\gamma^*$
\begin{eqnarray}
F_\pi(q^2) & = & f_{\pi\rho\omega}(q^2)
{1\over2}\{\frac{-m^2_\rho+i\sqrt{q^2}\Gamma_\rho(q^2)}{q^2-m^2_\rho+i\sqrt{q^2}
\Gamma_\rho(q^2)}+
\frac{-m^2_\omega+i\sqrt{q^2}\Gamma_\omega(q^2)}{q^2-m^2_\omega+i\sqrt{q^2}
\Gamma_\omega(q^2)}\},\nonumber \\
&&f_{\pi\rho\omega}(q^2)=
1+\frac{g^2}{2f^2_\pi}(1-{2c\over g})^2q^2.
\end{eqnarray}
$f_{\pi\rho\omega}$ is the intrinsic form factor of $\pi^0\gamma\gamma^*$.
For very low momentum we obtain
\begin{eqnarray}
F_\pi(q^2) & = & 1+a{q^2\over m^2_\pi},\\
&&a=
{m^2_\pi\over2}({1\over m^2_\rho}+{1\over m^2_\omega})
+{m^2_\pi\over2f^2_\pi}g^2(1-{2c\over g})^2.
\end{eqnarray}
The first term of Eq.(21) comes from the $\rho$ and $\omega$ poles
and the second term comes
from the intrinsic form factor of $\pi^0\gamma\gamma^*$.
The numerical value of a is
\begin{equation}
a=0.0303+0.0157=0.046.
\end{equation}
The first number of Eq.(22) is from the poles of vector mesons and the second number is from
the intrinsic form factor(19).
The contribution of the
poles of vector mesons is twice of the intrinsic form factor.
 
The decay amplitude(4) of $\pi^0\rightarrow\gamma\gamma$ is obtained
from ABJ anomaly in the chiral limit. It is interesting to see the
effect of the current
quark masses, $m_u$ and $m_d$, on the decay rate.
The current quark mass matrix is included in Eq.(1). Taking $m_{u,d}$
into the
calculation,
up to the first order in current quark masses we obtain
\begin{eqnarray}
{\cal L}_{\pi^0\rightarrow\gamma\gamma} & = & -\frac{\alpha}{\pi f_\pi}f_q
\pi^0\varepsilon^{\mu\nu\lambda\beta}
\partial_\mu A_\nu\partial_\lambda A_\beta,\\
&&f_q= 1-{1\over m}(m_u+m_d)+
{g^2\over2f^2_\pi}(1-{2c\over g})^2 m^2_{\pi^0},\\
&&\Gamma(\pi^0\rightarrow\gamma\gamma)=\frac{\alpha^2 m^3_\pi}{16\pi^3 f^2_\pi}
\{1-{1\over m}(m_u+m_d)+
{g^2\over2f^2_\pi}(1-{2c\over g})^2 m^2_{\pi^0}\}^2
\end{eqnarray}
$\frac{m_u+md}{m}$ needs to be determined. In Ref.[11] Eq.(1) has used to
predict all
the 10 Gasser-Leutwyler coefficients of the ChPT[12]. The masses of the
octet pseudoscalars
have been derived. Up to the second order in current quark masses they
are the
same as the ones obtained in ChPT. To the first order in $m_q$,
the pion mass is
expressed as[6,11]
\begin{equation}
m^2_\pi={4\over f^2_\pi}(-{1\over3})<\bar{\psi}\psi>(m_u+m_d),
\end{equation}
where $<\bar{\psi}\psi>$ is the quark condensate of three flavors.
In Ref.[11]
the quark condensate has been expressed as
\begin{equation}
-{1\over3}<\bar{\psi}\psi>=36g^4Qm^3,
\end{equation}
and it has been determined
\[Q=4.54.\]
It has been found in Ref.[11] that to the first order in current quark masses \(f_{\pi^0}=
f_{\pi^+}\).
From Eqs.(26,10,11) it is obtained
\begin{equation}
\frac{m_u+m_d}{m}={m^2_\pi\over f^2_\pi}{1\over4Q}(1-{2c\over g})^2.
\end{equation}
The quark mass factor(24) is expressed as
\begin{equation}
f_q=1+\frac{m^2_\pi}{f^2_\pi}(1-{2c\over g})^2\{-{1\over4Q}+{g^2\over2}\}
=1+4.97\times10^{-3}.
\end{equation}
The decay width of
$\pi^0\rightarrow\gamma\gamma$ is increased by $1\%$.
\begin{equation}
\Gamma_{\pi^0\rightarrow\gamma\gamma}=7.81\times f^2_q=7.89eV.
\end{equation}
The data is $7.83(1\pm0.071)eV$.
 
To conclude, it is found that in the form factor of $\pi^0\gamma\gamma^*$
besides the vector meson poles an additional
intrinsic form factor caused by quark loop is predicted. The slope of the form
factor of $\pi^0\rightarrow\gamma\gamma^*$ is predicted. The contribution of
the intrinsic form factor is about $50\%$ of the vector mesons. For
the form factor of $\pi^0\gamma\gamma^*$ at high $q^2$ all the derivative
should be taken into account. We will
present the study later. The effect of current quark masses on the decay rate
of $\pi^0\rightarrow\gamma\gamma$ is investigated. It is found that the rate
is increased by $1\%$.
 
The author wish to thank D.Dale and L.P.Gan for discussion. This study is supported
by a DOE grant.

\newpage
Figure Caption\\
Fig.1 Processes contributing to $\pi^0\gamma\gamma^*$

\begin{thebibliography}{30}
\bibitem{} S.L.Adler, Phys.Rev. {\bf 177},2426(1969); J.S.Bell and R.Jackiw,
Nuovo Cimento, {\bf 60A},47(1969).	
\bibitem{} H.J.Behrend et al., CLEO Coll., Z.Phys.,{\bf C49},401(1991);
D.M.Asner et al., CLEO Coll., CONF95-24, EPS0188,1995.
\bibitem{} R.M.Drees et al., Phys.Rev., {\bf D45},1439(1992); F.Farzanpay et al.,
Phy. Lett., {\bf B278},413(1992); H.Fonvieille et al., Phys.Lett.,
{\bf B233},651(1989);
P.Gumplinger, Thesis, Oregon State Univ.(1987); J.M. Poutissou et al.,
Proc. of Lake Louise Winter Inst.(1987), World Scientific, Singapore;
G.Tupper et al., Phys.Rev.{\bf D28},2905(1983); J.Fischer et al., Phys.Lett.,
{\bf 73B},359(1978); J.Burger Thesis, Columbia Univ.(1972); S.Devons et al.,
Phy.Rev.{\bf 184},1356(1969); N.P.Samios, Phys.Rev., {\bf 121},275(1961);
H.Kobrak, IL Nuovo Cimento, {\bf 20}, 1115(1961).
\bibitem{} A.Gasparian et al., The PrimEX Coll., proposal.
\bibitem{} L.G.Landsberg, Phys.Rep., {\bf 128},302(1985); J.Bijnes,A.Bramon
and F.Cornet,
Phys.Rev.Lett., {\bf 61},1453(1988); B.Moussallam, Phys.Rev.{\bf D51},4939(1995);
J.Bijnes, A.Bramon, F.Cornet, Z.Phys., {\bf C46},599(1990);P.Maris and
P.C.Tandy, Nucl.Phys.,{\bf A663},401c(2000); H.R.Frank,
Phys.Lett., {\bf B359},17(1995); H.Ito, W.Buck, F.Gross, Phys.Lett.,
{\bf B287},23(1992); Ivanov and Lyubovitskij, hep-ph/9705423.
\bibitem{} Bing An Li, Phys. {\bf D52},5165(1995); {\bf D52},5184(1995).
\bibitem{} G.'t Hooft, Nucl. Phys. {\bf B72}, 461(1974);{\bf B75}, 461(1974);
E.Witten, Nucl. Phys. {\bf B160}, 57(1979).
\bibitem{} J.Wess and B.Zumino, Phys.Lett., {\bf 37B},95(1971); E.Witten, Nucl.Phys.,
{\bf B223},422(1983).
\bibitem{} J.Gao and Bing An Li, Phys.Rev.{\bf D61}, 113006(2000).
\bibitem{} B.A. Li, Phys. Rev. {\bf D55}, 1436 (1997); {\bf D55} 1425 (1997);
D.N. Gao, B.A. Li, and M.L. Yan, Phys. Rev. {\bf D56}, 4115 (1997);
B.A. Li, D.N. Gao, and M.L. Yan, Phys. Rev. {\bf D58}, 094031
(1998);
\bibitem{} Bing An Li, hep-ph/9810311.
\bibitem{} J.Gasser anf H.Leutwyler,
Ann.Phys.(N.Y.){\bf 58},(1984)142;
Nucl.Phys. {\bf B250},(1985)465; {\bf B250}, (1985)517.
\end{thebibliography}
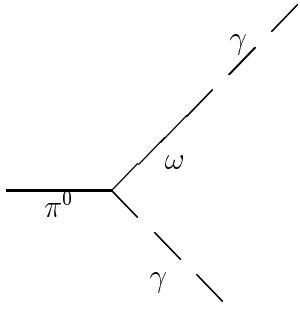
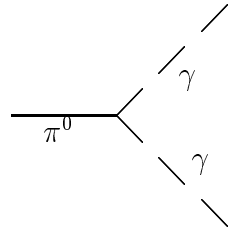
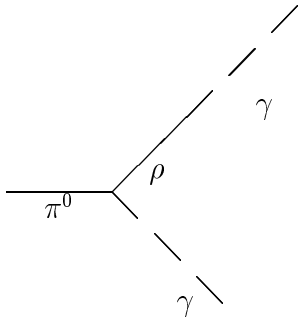
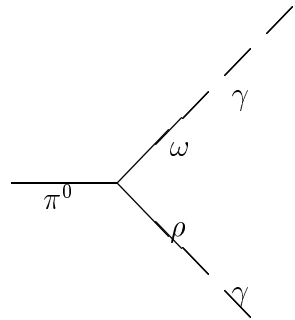
\end{document}